\documentclass[aps,prl,twocolumn,letterpaper,superscriptaddress,showpacs]{revtex4-1}
\usepackage[colorlinks,linkcolor=blue,anchorcolor=blue, citecolor=blue,urlcolor=blue,]{hyperref}
\usepackage{graphicx}
\usepackage{CJK}
\usepackage{verbatim}
\usepackage{physics}
\usepackage{ulem}

\begin{document}

\begin{CJK*}{UTF8}{bsmi}

\title{Terahertz nonlinear response in cuprate superconductors and the Higgs field in doped Mott insulators}
\author{Xiang Li (\CJKfamily{gbsn}李翔)}
\affiliation{Institute for Advanced Study, Tsinghua University, Beijing 100084, China}

\author{Zheng-Yu Weng}
\affiliation{Institute for Advanced Study, Tsinghua University, Beijing 100084, China}

\date{\today}

\begin{abstract}

A puzzling phenomenon has been recently revealed in terahertz nonlinear optical response experiments: A third harmonic generation (THG) signal, identified only in the superconducting (SC) phase in a conventional BCS superconductor, is found to persist into a wide pseudogap regime in the underdoped cuprate, accompanied by a $\pi$ phase shift in the THG signal across the SC transition. In this paper, we offer a consistent understanding of such an unconventional phenomenon based on an emergent Higgs mode of the condensed holons in the doped Mott insulator.  Specifically, in the lower pseudogap phase (LPP) where, although the holons still remain Bose condensed, the SC phase coherence gets disordered by excited spinons, which induce vortex-like responses from the holon condensate. By coupling to such internal fluctuations described by a mutual Chern-Simons gauge theory, we show that an external electromagnetic field can indeed produce the optical THG response in both the SC and LPP states, which are distinguished by a $\pi$ phase shift with a substantial suppression of the THG signal in the latter regime at higher temperatures.     
\end{abstract}

\maketitle
\end{CJK*}

\section{I. Introduction}

The mystery of the high-$T_c$ cuprate lies in its anomalous and complex responses to all kinds of experimental probes. Recent terahertz (THz) nonlinear optical measurements ~\cite{phaseshift_experiment,NanlinWang_experiments} in cuprates pose a new challenge with regard of how strongly correlated electrons respond to an optical pump by producing a clear nonlinear third harmonic generation (THG) signal beyond the superconducting state. Conventionally, a THG signal found in a superconductor described by the Bardeen-Cooper-Schrieffer (BCS) theory~\cite{BCStheory} is attributed to the existence of the Higgs (amplitude) modes of the Cooper pair condensation, which is only present below the superconducting (SC) transition temperature $T_c$ ~\cite{Review_HiggsinSC,Silaev}. In the literature, there has been an enormous effort to investigate a Higgs mode in condensed matter systems ~\cite{Review_HiggsinSC,Review_Higgs}. Surprisingly, the observed THG signals in the cuprate can even persist above $T_c$, indicating that if a Higgs mode is responsible for the nontrivial THG signal, it should still be present in a pseudogap region. In particular, a universal phase shift in the THG signal has been observed near $T_c$  ~\cite{phaseshift_experiment,NanlinWang_experiments}. 

These phenomena can be puzzling if one tries to analyze them within the BCS framework.
In the literature, early attempts include quasi-classical approaches \cite{Silaev,Comments_on_Silaev} and the gauge-invariant kinetic theory \cite{MW_Wu_kinetic_theory}.
The phase shift could be attributed to a resonance when the frequency of the THz pump coincides with the superconducting gap, $2\omega_0=2\Delta$~\cite{MW_Wu_kinetic_theory}.
However, this contradicts the non-zero THG signal observed above $T_c$, since the latter indicates that the gap is not closing near $T_c$.
Resonance was not observed in experiments \cite{phaseshift_experiment}.
In contrast, in Refs.~\cite{phaseshift_experiment,Fano_interference}, the THG signals above $T_c$ were phenomenologically attributed to the presence of preformed Cooper pairs in the pseudogap phase.
Moreover, the phase shift found near $T_c$ was interpreted in terms of the coupling of the Higgs modes with some other collective modes, such as charge density wave (CDW) fluctuations, rather than a resonance. Nevertheless, a satisfactory understanding supported by a microscopic framework based on the doped Mott insulator is still absent. Considering that the SC transition and the pseudogap regime in the cuprate are so complex and rich in phenomenon, an underlying self-consistent and unified framework is essential in order to fully understand the importance of the experiments in such a strongly correlated system.    

In this paper, we present a consistent description of the THG nonlinear response to a THz electromagnetic field observed in cuprates based on a new Higgs mode identified in the doped Mott insulator. Specifically, in this strongly correlated model, the electrons are fractionalized into the spinless holons and charge-neutral spinons due to the presence of a Mott gap to prevent the double occupancy of the electrons. Generally, it is expected that the holons can experience a Bose condensation while the spinons will form resonating-valence-bond (RVB) pairing condensation at a finite doping in the $t$-$J$ model \cite{Review_Lee_Nagaosa_Wen}.  However, there is generally an intrinsic long-range entanglement between the holon and spin degrees of freedom by the so-called phase string effect~\cite{Phasestring_PRL,Phasestring_general} in the $t$-$J$ model such that the two degrees of freedom are intrinsically coupled by mutual Chern-Simons (MCS) gauge fields~\cite{MCS}. Consequently, the holon (Higgs field) condensation does not necessarily correspond to the SC condensation as implied by the usual slave-boson mean-field theory (MFT) scheme \cite{Review_Lee_Nagaosa_Wen}. In fact, excited spinons can always induce a vortex-antivortex current response from the holon condensate via the MCS gauge fields to disorder the SC phase coherence above $T_c$, resulting in the so-called spontaneous vortex phase (SVP) or lower pseudogap phase (LPP) at finite temperatures even if the holons are still condensed. Such SC and SVP/LPP regimes are both characterized by the holon condensation, where the amplitude fluctuation of the holon condensate or the Higgs mode can emerge as a new excitation unique to such a doped Mott insulator system. The coupling between the Higgs mode and an external electromagnetic field can naturally lead to a distinct THG response in a nonlinear way in the SC phase below $T_c$ as well as the SVP/LPP above $T_c$. However, due to the phase disordering effect of the spinon-vortices in the SVP/LPP, the THz THG signals can get substantially reduced, together with a universal $\pi$ phase shift, which explains the experiments outlined in the beginning of this section.           

The remainder of the paper is organized as follows. In Section II, a phenomenological understanding of the THG experiments is outlined based on the present MCS theory. Here, an internal gauge fluctuation associated with vortices and antivortices induced by spin(on) excitations will modify the usual London equation for superfluid currents in a fundamental way, in both the SC and SVP/LPP states. 
In Section III, a microscopic effective theory based on the $t$-$J$ model is briefly outlined in Sec. A, where the phase string effect gives rise to the quantum long-range entanglement between the doped holes and background local moments via an MCS gauge structure. In particular, the holon condensation can further create a finite mass (gap) in the spin excitation via the MCS field. Otherwise, the latter would reduce to gapless spin wave in the antiferromagnetic (AFM) long-range order phase in the undoped case. Such a finite spin gap at finite doping has been previously shown \cite{Mei_TC} to further determine $T_c$ as well as the superfluid density \cite{Han_SCstiffness}, in a fashion similar to the Kosterlitz and Thouless (KT) transition ~\cite{KTtransition}. The underlying MCS gauge theory also give rise to a systematic understanding of the Nernst effect \cite{LPP,Phasestring_Nernst} and other transport properties \cite{Song_transport} in the SVP/LPP. In Sec. B of Section III, a systematic field-theoretical formulation of the MCS theory is presented, in which an effective Lagrangian description of the Higgs mode is obtained by integrating out all the other degrees of freedom of the matter fields and the MCS gauge fluctuations. Finally, in Sec. C, the THz THG response due to the coupling to the Higgs mode can be numerically calculated based on the derived effective theory. Section IV is devoted to the conclusion and discussion.   

\section{II. Phenomenological Understanding of the THG signal}
In this section, we shall present a phenomenological description of the THG signals based on a nonlinear coupling between the external electromagnetic (EM) field and the elementary excitations in the doped Mott insulator, whose microscopic theoretical treatment will be given in the next section. 

The conventional BCS superconductor and the present doped Mott insulator in the SC phase can be distinguished by the following Lagrangian density
\begin{eqnarray}
\label{Meissner}
\mathcal{L}_{\text{London}}= \left\{\begin{aligned}
&\frac{1}{2}\rho^{\text{bare}}_s(A^e_\alpha)^2, & \text{BCS}\\
&\frac{1}{2}\rho_h^{\text{bare}}(A^e_\alpha+A^s_\alpha)^2, &\text{P.S.}
\end{aligned}
\right.
\end{eqnarray}
where $\rho_s^{\text{bare}}$ is the bare superfluid density for a BCS superconductor and $\rho_h^{\text{bare}}$ denotes that in the phase-string theory (P.S.) for the doped Mott insulator.
$\alpha$ (and also $\beta$ in the rest of this paper) denotes the 2D spatial components, $x$ and $y$.
The corresponding London equations~\cite{Londonequation} are given by
\begin{eqnarray}
\label{LondonEquation}
J^{\alpha} \equiv \frac{\delta \mathcal{L}_{\text{London}}}{\delta A_\alpha^e}
=\left\{\begin{aligned}
&\rho^{\text{bare}}_s(A^{e,\alpha}), & \text{BCS}\\
&\rho_h^{\text{bare}}(A^{e,\alpha}+A^{s,\alpha}). &\text{P.S.}
\end{aligned}
\right.
\end{eqnarray}
The dynamical internal gauge field $A^s_\alpha$ in the second line of Eq. (\ref{LondonEquation}) (P.S.) characterizes the excited vortices with its flux density proportional to the vortex density $n_v$ by 
\begin{eqnarray}
\label{vortex_and_As}
\begin{aligned}
     n_v=\frac{1}{\pi}\epsilon^{\alpha\beta}\partial_\alpha A^s_\beta~.
\end{aligned}
\end{eqnarray}
The proportionality constant in Eq.(\ref{vortex_and_As}) is $1/\pi$ rather than $1/2\pi$, which corresponds to $\pi$-vortices as perceived by the charge $+e$ holon condensate, where the bosonic holon field can be regarded as the Higgs field in a fractionalization formulation of the doped Mott insulator based on the phase-string formalism of the $t$-$J$ model~\cite{Phasestring_general,Phasestring_meanfield}. 

Therefore, the basic starting point of the present phenomenological theory for the SC phase in the underdoped cuprate is dictated by a topological phase transition at $T_c$ above which vortex-like excitations proliferate in a similar fashion as in the KT transition ~\cite{KTtransition}. Nevertheless, different from the conventional KT vortices, the present vortex excitations, whose density is denoted by $n_v$, are quantum-like with a charge-neutral spin-1/2 (spinon) sitting at the core of each vortex. In this way, at low temperatures, the vortex and antivortex are paired up but do not annihilate each other due to the presence of spinons at the cores which remain in a spin-singlet RVB pairing in the ground state. The minimal energy cost of creating a pair of free spinon-vortex excitations is thus ``cheap vortices'' without involving the creation of vortices from a vacuum. It is essentially the energy cost of breaking up an RVB pair with a minimal energy $E_g$ (see Sec. III), and the SC phase transition caused by the proliferation of the spinon-vortices is then determined by $E_g$ instead of the conventional core energy in the classical KT transition. Above $T_c$, the proliferating spinon-vortices will dominate the SC flucutations in the phase called SVP or LPP, which is characterized by anomalous transports including the Nernst effect.

In a conventional BCS superconductor, the optical nonlinear THG signal primarily probes the coupling between the fluctuation of the Cooper pair amplitude and the external EM vector potential $A^e_\alpha$ ~\cite{Review_HiggsinSC}. A straightforward generalization according to Eq. (\ref{Meissner}) gives rise to the following term in the Lagrangian density
\begin{eqnarray}
\label{THGcoupling}
\mathcal{L}_{\text{THG}}= \left\{\begin{aligned}
&\frac{1}{2}k_{\text{THG}}H(A^e_\alpha)^2, & \text{BCS}\\
&\frac{1}{2}k_{\text{THG}}H_h(A^e_\alpha+A^s_\alpha)^2, &\text{P.S.}
\end{aligned}
\right.
\end{eqnarray}
where $H$ denotes the amplitude fluctuation of the Cooper pair condensate and $H_h$ the amplitude fluctuation (the Higgs mode) of the holon condensate, and $k_{\text{THG}}$ is the coupling constant. Then the Higgs-mode-driven nonlinear current responses are given, respectively, by  
\begin{eqnarray}
\label{THGcuurent}
J^\alpha_{\mathrm{THG}}\equiv \frac{\delta \mathcal{L}_{\text{THG}}}{\delta A^e_\alpha} \propto \left\{\begin{aligned}
&HA^{e,\alpha}, & \text{BCS}\\
&H_h(A^{e,\alpha}+A^{s,\alpha}). &\text{P.S.}\\
\end{aligned}
\right. 
\end{eqnarray}
Since the Higgs modes are driven by $(A^e_\alpha)^2$ or $(A^e_\alpha+A^s_\alpha)^2$, respectively, in Eq. (\ref{THGcoupling}), the equations of motion of the Higgs modes give rise to
\begin{eqnarray}
\label{HiggsEOM}
\begin{aligned}
& H\propto (A^e_\alpha)^2, & \text{BCS}\\
& H_h\propto (A^e_\alpha+A^s_\alpha)^2, &\text{P.S.}
\end{aligned}
\end{eqnarray}
Substituting Eq.~(\ref{HiggsEOM}) to Eq.(\ref{THGcuurent}), we get
\begin{eqnarray}
\label{THGcuurent_2}
J^\alpha_{\mathrm{THG}}\propto \left\{\begin{aligned}
&(A^{e,\alpha})^3, & \text{BCS}\\
&(A^{e,\alpha}+A^{s,\alpha})^3. &\text{P.S.}\\
\end{aligned}
\right. 
\end{eqnarray}

\begin{figure}
	 \setlength{\belowcaptionskip}{-0.5cm}
	\vspace{-0.1cm}
	\begin{center}
		\includegraphics[width=0.9\columnwidth]{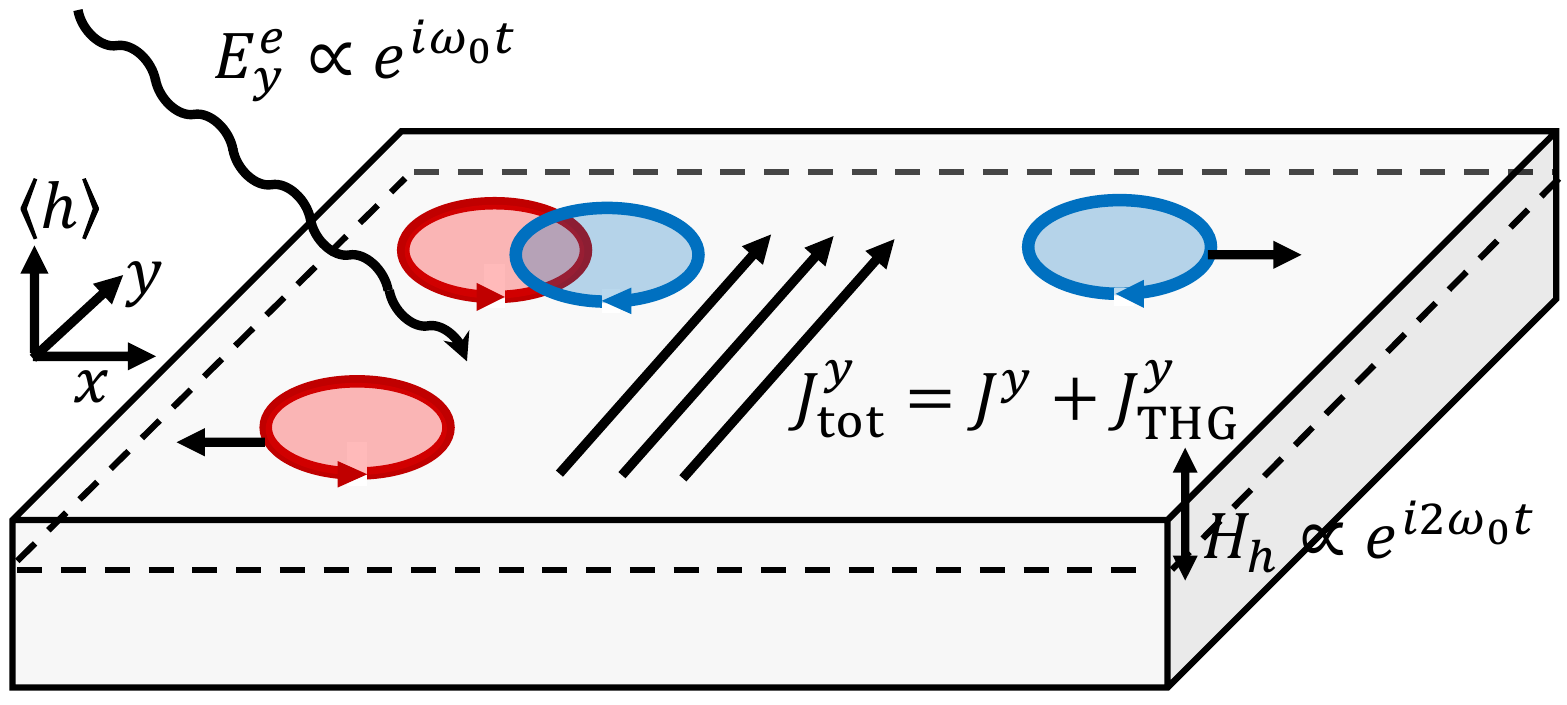}
	\caption{Response of the charged condensate to an applied electric field. 
    The vertical direction represents the condensation amplitude $\langle h\rangle$. 
    The electric field is along $y$ direction in this demonstration.
    In addition to the linear response current $J^y=\rho^\text{bare}_h(A^{e,y}+A^{s,y})$, there is also a non-linear current $J_\text{THG}^y=k_\text{THG}H_h(A^{e,y}+A^{s,y})$ due to the amplitude fluctuation of $\rho^\text{bare}_h$, i.e., the Higgs mode denoted by $H_h$.
    The Higgs mode is coupled to vortex-like excitations, which thus have influence on the THG current.
    The vortices are in red and the antivortices are in blue, and the circulating arrows represent the supercurrents around them..
    The vortex-like excitations are confined and form vortex-antivortex pair below $T_c$, as shown in the top left of the figure.
    They are deconfined and able to move freely above $T_c$, as shown in the top right and the bottom left.
    These different behaviors above and below $T_c$ will lead to the phase shift of the THG signal, as discussed in text.}
	\label{fig_Higgs_Mode}
	\end{center}
\end{figure}

In the SC phase, under an oscillating EM field $A^e_\alpha \propto e^{i\omega_0 t}$ with frequency $\omega_0$, one finds $J^\alpha_{\mathrm{THG}}\propto e^{i3\omega_0 t}$ for the BCS case, which gives rise to a THG signal at $T<T_c$. However, due to the vanishing Cooper pairing, the THG disappears in general above $T_c$ in the BCS theory. By contrast, the bare holon condensation will remain finite at $T>T_c$ in the SVP (LPP) of the phase-string theory, which sustains the excitations of the Higgs mode $H_h$ such that Eq. (\ref{THGcuurent_2}) is still valid in such a regime above $T_c$. We shall show that $ A^{s}_\alpha \propto A^{e}_\alpha$
based on the microscopic MCS theory in the next section, and thus $J^\alpha_{\mathrm{THG}}\propto e^{i3\omega_0 t}$ holds true in both SC and SVP (LPP) regimes.
The Higgs mode and the currents are demonstrated schematically in Fig.~\ref{fig_Higgs_Mode}.

In particular, in the SC phase, the London equation remains true: $J^{\alpha}=
\tilde{\rho}_h A^{e,\alpha}$ such that 
\begin{equation}
\label{T<Tc}
A^e_\alpha+A^{s}_\alpha \propto A^{e}_\alpha,\;\;\;\; T<T_c,
\end{equation}
according to Eq.(\ref{LondonEquation}) for the P.S. case, where $\tilde{\rho}_h$ denotes the renormalized superfluid density of the holon condensate. Here, due to the confinement of the vortex-antivortex pairs, one may take a proper gauge such that $A^s_{\alpha}\rightarrow 0$ in the absence of $A^e_{\alpha}$. But an external $A^e_{\alpha}$ can induce a nontrivial response of $A^s_{\alpha}\propto A^{e}_\alpha$ such that the bare superfluid density $\rho_h^{\text{bare}}$ always gets renormalized into $\tilde{\rho}_h$.

\begin{figure}
\setlength{\belowcaptionskip}{-0.5cm}
	\vspace{-0.1cm}
	\begin{center}
		\includegraphics[width=\columnwidth]{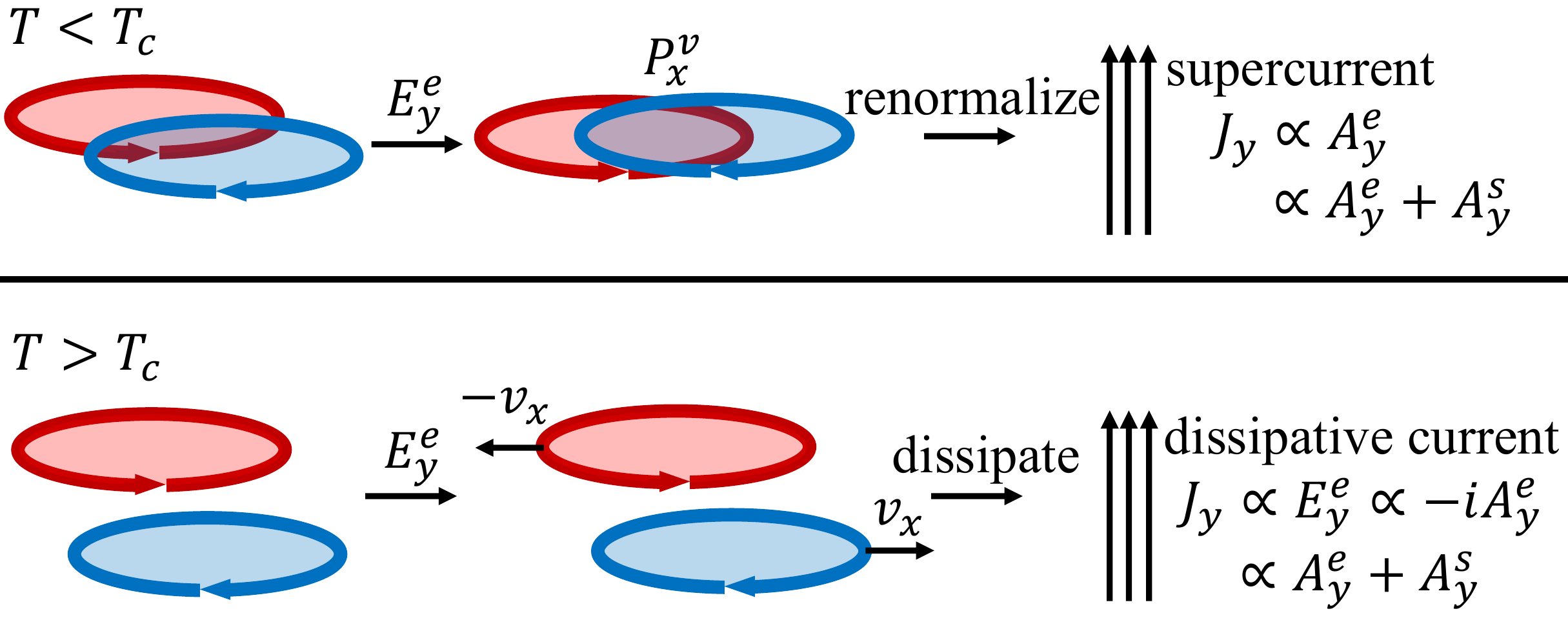}
	\caption{The response of spin-vortex to electric field. Vortex is represented in red and antivortex is in blue. The circulating arrows indicate the direction of the supercurrents of the charged condensate. Suppose the electric field $E^e_y$ is along $y$ direction. Below $T_c$, the confined vortex-antivortex pairs are aligned along $x$ direction by $E^e_y$, leading to a vorticity polarization $P_x^v$, which is viewed as a response gauge field $A^s_y$ that renormalizes the supercurrent. Above $T_c$, the spinon-vortices are deconfined and can thus move freely. A driven force $E^e_y$ will cause an averaged vortex motion along $x$ direction, which bring dissipation into the electric transport. The phase shift of $A^e_\alpha+A^s_\alpha$ can be understood from the change of nature of the first harmonic $(1\omega)$ current across $T_c$.}
	\label{fig_pi_shift}
	\end{center}
\end{figure}

In the SVP/LPP regime, the SC phase coherence is disordered by the proliferation of free spinon-vortices via $A^s_{\alpha}$. In such a ``normal state'' at $T>T_c$, a finite optical conductivity $\sigma(\omega)$ gives rise to 
\begin{eqnarray}
\label{opticalconductivity}
\begin{aligned}
J^\alpha(\omega)=\sigma (\omega)E^{e,{\alpha}}(\omega),
\end{aligned}
\end{eqnarray}
which is approximately a real number (the value of the Drude peak)\cite{opticalconductivity} at small frequency (THz). Then by combining with Eq.~(\ref{LondonEquation}), with choosing a proper gauge: $E^{e}_{\alpha}=-\partial_tA^e_\alpha\rightarrow E^e_\alpha(\omega)=-i\omega A^e_\alpha(\omega)$,
one finds
\begin{eqnarray}
\label{SVPAArelation}
\begin{aligned}
\label{T>Tc}
A^e_\alpha+A^{s}_\alpha \propto -iA^{e}_\alpha,\;\;\;\; T>T_c,
\end{aligned}
\end{eqnarray}  
which is in contrast with $A^e_\alpha+A^{s}_\alpha \propto A^{e}_\alpha$ at $T<T_c$ by a phase shift $i$. The arguments are demonstrated in Fig.~\ref{fig_pi_shift}.

Define the phase of the THG signal by
\begin{eqnarray}
\label{phaseofTHG}
\begin{aligned}
e^{i\Theta} \equiv \frac{J^{\alpha}_\text{THG}(3\omega_0)}{J^{\alpha} (\omega_0)}\cdot\left\vert\frac{J^{\alpha}_\text{THG}(3\omega_0)}{J^{\alpha}(\omega_0)}\right\vert =\frac{(A^e_{\alpha}+A^s_{\alpha})^2}{\left\vert A^e_{\alpha}+A^s_{\alpha}\right\vert^2}
\end{aligned}
\end{eqnarray}
The phase shift across $T_c$ is then determined by
\begin{eqnarray}
\label{phaseshift}
\begin{aligned}
\Delta\Theta\equiv\Theta(T<T_c)-\Theta(T>T_c)
\end{aligned}
\end{eqnarray}
which gives rise to $\Delta\Theta=2\times\frac{\pi}{2}=\pi$.
Such a total $\pi$ phase shift across $T_c$ agrees with the THG experimental measurement, which is universal without involving detailed dynamical properties in the reasoning.

We finally remark that, in general, Eqs. (\ref{T<Tc}) and (\ref{T>Tc}) can be combined in the following formula for both SC and SVP/LPP based on the microscopic theory to be given in the next section:
\begin{eqnarray}
\label{Ae+As}
\begin{aligned}
    A^e_\alpha+A^s_\alpha&=\frac{1}{1+\pi^2\sigma^h\sigma^s}A^e_\alpha\\
    &=\frac{\omega^2}{\omega^2-\pi^2 \rho_h^{\text{bare}}K(\omega)}A^e_\alpha,
\end{aligned}
\end{eqnarray}
where $\sigma^h$ is the optical conductivity of the holons: $\sigma^h=i\frac{\rho_h^{\text{bare}}}{\omega}$ and $\sigma^s$ is the conductivity of the spinons $\sigma^s\equiv i\frac{K(\omega)}{\omega}$, where $K$ is the longitudinal current-current correlation (with both ``paramagnetic" contribution and ``diamagnetic" contribution) of spinons in the long-wavelength limit ($\mathbf{p}=0$). In the SC phase where the spinons are ``confined'' (see Sec. III), one has $K(\omega)\approx-\chi^s_e\omega^2+O(\omega^3)$ at $\omega=\omega_0 \sim \text{THz}$, such that Eq. (\ref{T<Tc}) is recovered, where $\chi^s_e$ is the ``electric" susceptibility for the spinon-vortices. In the SVP/LPP state above $T_c$, the deconfined spinons give rise to $K(\omega)\approx-\chi_e^s\omega^2-i\sigma^s_0\omega+O(\omega^3)$, where $\sigma^s_0$ denotes the DC conductivity of the spinons. 
In this case, Eq.~(\ref{Ae+As}) is reduced to Eq.~(\ref{T>Tc}), provided that $\omega<<\pi^2 \rho_h^{\text{bare}}\sigma^s_0$ presumably satisfied by the THz frequency $\omega_0$. 

\section{III. Microscopic Description}
Having offered a phenomenological description of the THz THG experiment, we now turn to justifying it based on a microscopic theoretical description of the $t$-$J$ model. In a doped Mott insulator, charges (doped holes) are injected into a Mott insulator of the half-filled electron system with a charge (Mott) gap, which prevents double occupancy of the electrons in the strong coupling regime described by the $t$-$J$ model. Here, the relevant degrees of freedom can be separated into local moments and doped charges (holes) and distinct theoretical descriptions concern how these elementary excitations are characterized. For example, in the slave-boson mean-field theory framework, they are described by fermionic neutral ($s=1/2$) spinons and spinless bosonic holons coupled by U(1) or SU(2) gauge fields \cite{Review_Lee_Nagaosa_Wen}. In the phase-string formulation, on the other hand, both the spinons and holons are bosons with a mutual Chern-Simons topological gauge structure due to the phase-string effect hidden in the $t$-$J$ model (see below). In the following, we focus on the latter theory and show how the THG signal provides an interesting experimental probe into the two low-temperature phases, where the Higgs mode involving the holon condensate and the gauge fluctuations at the SC phase coherence transition play an essential role. 

\subsection{A. Phase-string-theory description of the $t$-$J$ model  }
We start with the $t$-$J$ model on a two-dimensional square lattice:
\begin{eqnarray}
\label{tJmodel}
\begin{aligned}
&H_{t\text{-}J}=H_t+H_J\\
&\equiv-t\sum_{\langle i,j\rangle,\sigma}c^\dagger_{i\sigma}c_{j\sigma}+h.c.+J\sum_{\langle i,j\rangle}\left(\mathbf{S}_i\cdot\mathbf{S}_j-\frac{n_in_j}{4}\right),
\end{aligned}
\end{eqnarray}
with the no-double-occupancy constraint $\sum_\sigma c^\dagger_{i\sigma}c_{i\sigma}\leq1$. This inequality constraint can be further replaced by an equality constraint via fractionalizing the electron creation and annihilation operator $c^\dagger$ and $c$ \cite{Review_Lee_Nagaosa_Wen}. 
In the phase-string formulation, the electron can be fractionalized into three types of partons~\cite{Phasestring_review}: 1) the holons created by $h^\dagger$, which are spinless bosons; 2) the spinons created by $b^\dagger_\sigma$, which are charge-neutral bosons; and 3) the backflow spinons created by $a^\dagger_\sigma$, which are charge-neutral fermions that represents the spin $\frac{1}{2}$ associated with the doped holes. (With the introduction of the backflow spinons, $a$, the bosonic spinons, $b$, are kept always at half-filled with one spinon per lattice site~\cite{Phasestring_review}.) Such an electron fractionalization can naturally incorporate the phase-string effect as a singular topological Berry phase hidden in the $t$-$J$ model ~\cite{Phasestring_PRL,Phasestring_general,Phasestring_signstructure} via the emergent gauge fields, $A^h_\mu$ and $A^s_\mu$ in the following low-energy effective Lagrangian $L=L_h+L_b+L_\text{MCS}$:
\begin{eqnarray}
\label{Lagrangian_phasestring}
\begin{aligned}
L_h = &\sum_I h^{\dagger}_I(\partial_\tau-iA_0^s-iA^e_0)h_I\\
&-t_h\sum_{I,\alpha}(h^{\dagger}_{I+\alpha}h_I e^{iA_\alpha^s(i)+iA_\alpha^e} +h.c.),\\
L_b = &\sum_{i,\sigma}b^{\dagger}_{i,\sigma}(\partial_\tau-i\sigma A^h_0+\lambda_b)b_{i,\sigma}\\
&-\frac{J_{\text{eff}}}{2}\Delta_s\sum_{i,\alpha,\sigma}(b^{\dagger}_{i+\alpha,\sigma}b^{\dagger}_{i,\bar{\sigma}}e^{i\sigma A_\alpha^h(i)}+h.c.),\\
L_\text{MCS} &= \frac{i}{\pi}\sum_i\epsilon^{\mu\nu\lambda}A^s_\mu(I)\partial_\nu A^h_\lambda(i).
\end{aligned}
\end{eqnarray}
Here and in the rest of this paper, $\mu,\nu,\lambda=0,1,2$ denote the $\tau,x,y$ components, respectively. 
$i$ labels vertices on a two-dimensional square lattice, and $I$ labels dual lattice sites~\cite{MCS}.
The holons (the $h$ field) in the Lagrangian $L_h$ are coupled to the electromagnetic field $A^e_\mu$, and the spinons (the $b$ field) in the Lagrangian $L_b$ form bosonic RVB pairing ~\cite{Anderson_RVB}, characterized by the order parameter $\Delta_s=\langle\sum_{\sigma}b^{\dagger}_{i+\alpha,\sigma}b^{\dagger}_{i,\bar{\sigma}}e^{i\sigma A_\alpha^h(i)}\rangle$ with the Lagrangian multiplier $\lambda_b$ enforcing one $b$-spinon per site on average.
The holon and spinon fields are further coupled to the U(1) gauge fields,  $A^s$ and $A^h$, in $L_h$ and $L_b$, respectively, which are mutually connected by the mutual Chern-Simons term $L_\text{MCS}$ in Eq. (\ref{Lagrangian_phasestring}). Here, $L_\text{MCS}$ characterizes a mutual entanglement between holons and spinons as shown by the following equations of motion:
\begin{eqnarray}
\label{fluxattachment}
\begin{aligned}
j^\mu_s &= \frac{-i}{\pi}\epsilon^{\mu\nu\lambda}\partial_\nu A_\lambda^s,\\
j^\mu_h &= \frac{-i}{\pi}\epsilon^{\mu\nu\lambda}\partial_\nu A_\lambda^h,
\end{aligned}
\end{eqnarray}
where the spinon current $j^\mu_s$ and the holon current $j^\mu_h$ are defined by $j^\mu_s = \frac{\partial L_b}{\partial A_\mu^h}$ and $j^\mu_h = \frac{\partial L_h}{\partial A_\mu^s}$, respectively. In particular, the density components of $\mu=0$ explicitly indicate mutual $\pi$-flux attachments or mutual semion statistics between the $h$ holons and $b$ spinons \cite{Phasestring_review}.
It is noted that the backflow fermionic $a$ spinons remain tightly paired \cite{Phasestring_Ma_2014}, which are of high energy and not presented in Eq.~(\ref{Lagrangian_phasestring}) since the low-energy/low-temperature physics is mainly concerned in this work. The hopping integral $t_h$ and superexchange coupling $J_\text{eff}$ are the renormalized parameters at a generalized mean-field level \cite{Phasestring_Ma_2014}.

\begin{figure}
	 \setlength{\belowcaptionskip}{-0.5cm}
	\vspace{-0.1cm}
	\begin{center}
		\includegraphics[width=8cm]{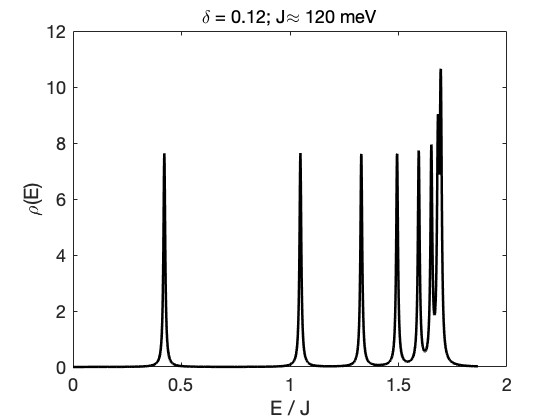}
	\caption{The spectrum of spinon shown by its density of states. The spinon excitations have a finite gap, which is resulted from the holon condensate. In this sense, not only the photons but also the spinons acquire mass by the Higgs mechanism. A broadening in this figure is introduced to demonstrate the Dirac delta functions in the density of state.}
	\label{fig_spin_DOS}
	\end{center}
\end{figure}

At half-filling, the holon number is equal to zero. Here, one may drop both $L_h$ and $L_\text{MCS}$ in Eq. (\ref{Lagrangian_phasestring}) and set $A^h=0$ in $L_b$, which is simply reduced to that of the Schwinger-boson mean-field description of the antiferromagnetic ground state~\cite{Auerbach_SchwingerBosonMF}. At finite doping, new low-temperature phases emerge, characterized by the Bose condensation of holons with $\langle h_I \rangle=\sqrt{\rho_h^0} e^{i\theta_I}$ in which $\rho_h^0\propto \delta$, the doping level. Consequently, a finite average flux [i.e., $\delta\pi$ per plaquette according to the $\mu=0$ component in the second line of Eq.~(\ref{fluxattachment})] will be described by $A^h$, which exerts on the $b$-spinons via $L_b$, leading to a solution of an RVB state with short-range antiferromagnetic correlation and a spin gap opened up as shown in the spinon spectrum in Fig.~\ref{fig_spin_DOS}. Namely, the holons as a ``Higgs field'' gives rise to a ``mass'' (gap) to the spinons.   

The corresponding $L_h$ in Eq. (\ref{Lagrangian_phasestring}) becomes  
\begin{equation}\label{LG}
L_h \rightarrow -2t_h\rho_h^0\sum_{I,\alpha}\cos{(\partial_{\alpha}\theta-A^e_\alpha-A^s_\alpha)}
\end{equation}
with choosing a proper gauge by setting the temporal component zero. This effective Lagrangian in Eq.~(\ref{LG}) will characterize both the SC and the SVP/LPP states ~\cite{GeneralizedGLtheory,LPP}. Here, the internal gauge field $A^s_\alpha$ describes the fictitious $\pi$-fluxes attached to $b$-spinons according to Eqs.~(\ref{fluxattachment}) and (\ref{vortex_and_As}), which will induce a $\pm \pi$ vortex supercurrent response according to Eq. (\ref{LG}) for each single spinon. The latter sits at the vortex core to form a spinon-vortex composite ~\cite{GeneralizedGLtheory,LPP}. Note that in a short-range RVB background, $A^s_\alpha$ can be effectively cancelled or ``screened'' by the RVB pairing. In the SC phase at low temperature below $T_c$, $A^s_\alpha$ remains to be ``screened'' as the $b$-spinons excited from the RVB background are ``confined'' in vortex-antivortex bound pairs, known as the spin-rotons ~\cite{Mei_TC,GeneralizedGLtheory,LPP} such that $L_h$ in Eq. (\ref{LG}) gives rise to a London-like charge current equation (cf. Sec. II) for the SC phase. Such a superconducting phase is thus similar to the quasi-long-range ordered phase of the 2D XY model studied by Kosterlitz and Thouless~\cite{KTtransition}, except that each spinon-vortex has a vorticity $\pi$ rather than $2\pi$, and in particular, each vortex core is associated with a quantum spin-1/2 spinon. It has been previously shown \cite{Mei_TC} that the SC phase transition temperature $T_c$ can be essentially determined by a minimal gap $E_g$ of the spin-roton excitation. 

Here, one may also understand the SC phase transition in a dual view (the Coulomb gas representation) by using the 2D boson-vortex duality~\cite{CoulombGas}, where
spinon-vortices can be viewed as ``charged" particles. The vorticity $v=\pm1$ is viewed as positive or negative unit ``charge", while the logarithmic confining potential is precisely the 2D Coulomb potential. The SC phase is an ``insulating state'' composed of ``electric" dipoles of confined spinon-vortex pairs on top of the spin-singlet-pairing RVB background. The SVP/LPP regime above $T_c$ is composed of the plasma of ``electric charges'' as a spinon-vortex conducting phase.

We have seen that the $b$ spinon excitations are vortex-like excitations in the SC and LPP regime, which are called spinon-vortex composites~\cite{Mei_TC}. A spinon-vortex is different from an ordinary vortex in a classical KT system in two aspects:
1) There is a spin-$\frac{1}{2}$ at the vortex core; 2) According to Eq.~(\ref{fluxattachment}), the spin $\uparrow$ or $\downarrow$ spinons are viewed as $+\pi$ or $-\pi$, rather than $\pm 2\pi$, vortices, respectively~\cite{Mei_TC}. Such $\pm\pi$ spinon-vortices can fuse with ordinary $\mp 2\pi$ vortices to become composite excitations with $\mp 1$ vorticity on the square lattice. (Here the vorticity is defined modulo $\pi$, rather than $2\pi$.)
Thus, there are totally four types of spinon-vortex excitations, labeled by spin $\sigma$ and vorticity $v$, which are of the same absolute vorticity and thus degenerate in energy~\cite{Mei_TC}.
So, one needs to consider all of them while ignoring other excitations with higher vorticity at low temperature.
Denoting spinons as $b_{\sigma=+1,v=+1}$ and $b_{\sigma=-1,v=-1}$, and composite excitations as $b_{\sigma=+1,v=-1}$ and $b_{\sigma=-1,v=+1}$, the spinon compact Lagrangian $L_b$ in Eq.~(\ref{Lagrangian_phasestring}) can be reexpressed by
\begin{eqnarray}
\label{Lb}
\begin{aligned}
L_b \rightarrow &\sum_{i,\sigma,v}b^{\dagger}_{i,\sigma,v}(\partial_\tau-iv A^h_0+\lambda_b)b_{i,\sigma,v}\\
&-\frac{J_{\text{eff}}}{2}\Delta_s\sum_{i,\alpha,\sigma,v}(b^{\dagger}_{i+\alpha,\sigma,v}b^{\dagger}_{i,\bar{\sigma},\bar{v}}e^{iv A_\alpha^h(i)}+h.c.)~,
\end{aligned}
\end{eqnarray}
in which the compactness of the original $L_b$ may be relaxed in taking a continuum limit later in the next subsection.

Therefore, the distinction between the SC phase and the SVP/LPP is essentially dictated by the fluctuation of the internal gauge field $A^s_\alpha$, while the holon condensation remains in both phases. In the following, we briefly provide an intuitive picture to show how $A^s_\alpha$ responds linearly to the external $A^e_\alpha$ differently below and above $T_c$ based on the above mutual Chern-Simons gauge theory formulation. According to Eq.~(\ref{fluxattachment}), the holon current can be produced by a total ``electric" field $E^h_\alpha \equiv-\partial_\alpha A^h_t-\partial_tA^h_\alpha$, which linearly induces a spinon current $j^\alpha_s$.
\begin{eqnarray}
\label{Aeffderivation}
\begin{aligned}
    J^\alpha = j^{\alpha}_h&=\sigma^h(E^{s,\alpha}+E^{e,\alpha})\\ 
    &=\frac{1}{\pi}\epsilon^{\alpha\beta}E^{h}_\beta\\
    &=\frac{1}{\pi}\epsilon^{\alpha\beta}\frac{1}{\sigma^s}j_{s,\beta}\\
    &=\frac{1}{\pi\sigma^s}\epsilon^{\alpha\beta}(\frac{1}{\pi}\epsilon_{\beta\rho}E^{s,\rho}).\\
    &=-\frac{1}{\pi^2\sigma^s}E^{s,\alpha}.
\end{aligned}
\end{eqnarray}
In the first line, we used the definition of the optical conductivity of the holons $\sigma^h=i\frac{\rho_h^{\text{bare}}}{\omega}$.
The second line and the fourth line follows from Eq.~(\ref{fluxattachment}).
We used the definition of the ``optical" conductivity of the spinons $\sigma^s\equiv i\frac{K(\omega)}{\omega}$, where $K$ is the longitudinal current-current correlation (wiht both ``paramagnetic" contribution and ``diamagnetic" contribution) of spinons in the long-wavelength limit ($\mathbf{p}=0$), which can be calculated directly from the Lagrangian Eq.~(\ref{Lb}).
We find 
\begin{eqnarray}
\label{Aeffresult}
\begin{aligned}
    E^e_\alpha+E^s_\alpha&=\frac{1}{1+\pi^2\sigma^h\sigma^s}E^e_\alpha,
\end{aligned}
\end{eqnarray}
which leads to Eq. (\ref{Ae+As}) in Section II.

In this work, we shall further explore a new elementary excitation, i.e., the Higgs mode $H_h$, associated with the amplitude fluctuation ~\cite{Review_HiggsinSC,Review_Higgs} of the holon condensate:
\begin{equation}\label{Hh}
h_I=\left(\sqrt{\rho_h^0}+H_h\right) e^{i\theta_I}~.
\end{equation}
It is noted that the usual Nambu-Goldstone mode corresponds to the phase fluctuation $\theta_I$ in the condensate. 
Minimally coupled to the electromagnetic field, the Nambe-Goldstone mode will become a high energy plasmon mode according to the Anderson-Higgs mechanism~\cite{Review_HiggsinSC,Anderson_Higgs,HiggsMechanism}, which may be ignored in the present work. The remaining Higgs mode in Eq.(\ref{Hh}) will be present both below and above $T_c$ so long as the Higgs field (holon) is condensed. As phenomenologically outlined in the last section, an external EM field coupled with $H_h$ will produce an unconventional THG signal, which can persist over to the SVP/LPP regime with $\rho_h^0\neq 0$. 

\subsection{B. Field-Theoretical Formulation}
Mathematically, the effective action governing the dynamics of the Higgs mode $H_h$ can be obtained by integrating out all the other degrees of freedom in the Lagrangian (\ref{Lagrangian_phasestring}) as follows: 
\begin{eqnarray}
\begin{aligned}
Z&=\int \mathcal{D}h\mathcal{D}A^s\mathcal{D}A^h\mathcal{D}b\;e^{-\int d\tau L[h,A^s,A^h,b]}\\  
&=\int \mathcal{D}h\mathcal{D}A^s\mathcal{D}A^h\;e^{-\int d\tau L_h+L_\text{MCS}}\int\mathcal{D}b\;e^{-\int d\tau L_b}\\ 
&\approx\int \mathcal{D}h\mathcal{D}A^se^{-\int d\tau L^h}\int\mathcal{D}(\Delta A^h)e^{-\int d\tau L_\text{MCS}+L^{\text{eff}}_{\Delta A^h}}\\ 
&=\int \mathcal{D}h\int\mathcal{D}A^s \;e^{-\int d\tau L_h+L^{\text{eff}}_{A^s}}\\ 
&\approx\int\mathcal{D}H_h\mathcal{D}\theta\int\mathcal{D}A^s \;e^{-\int d\tau L_h[H_h,\theta,A^s;A^e]+L^{\text{eff}}_{A^s}}\\ 
&=\int\mathcal{D}H_h\int\mathcal{D}A^s \;e^{-\int d\tau L_h[H_h,A^s;A^e]+L^{\text{eff}}_{A^s}} \\ 
&=\int \mathcal{D}H_h\;e^{-\int d\tau L^\text{eff}_{H_h}[H_h,A^e]}~.
\label{integratingout}
\end{aligned}
\end{eqnarray}
Here, the first step is to integrate out the spinon field $b$ in the second line and obtain a functional of the fluctuation $\Delta A^h\equiv A^h-\bar{A}^h$ in the third line ($\bar{A}^h$ is to be defined below).
After integrating out the fluctuating gauge field $\Delta A^h$, one can obtain a new term as a functional of $A^s$ in the fourth line;  In the holon condensed phases, the holon (Higgs) field may be decomposed as in Eq. (\ref{Hh}) in fifth line;
Then by taking a unitary gauge to absorb the Goldstone mode associated with the $\theta$ field [cf. Eq. (\ref{LG})] in the sixth line, one can
finally integrate out the gauge field $A^s$ to get an effective Lagrangian for the Higgs mode in the last line.

We begin with integrating out spinons $b$ to get an effective Lagrangian $L^\text{eff}_{\Delta A^h}$ in the second and third lines in Eq.~(\ref{integratingout}). Define $\Delta A^h_\alpha\equiv A^h_\alpha-\bar{A}^h_\alpha$ and $\Delta A^h_0\equiv A^h_0$, where the averaged $\bar{A}^h$ is defined by $\epsilon^{\alpha\beta}\partial_\alpha \bar{A}^h_\beta\equiv \delta\pi$, that is, the averaged field strength of $\bar{A}^h_\alpha$ is attached to the averaged holon density $\delta$ according to Eq.~(\ref{fluxattachment}). Assuming the gauge fluctuation of $\Delta A^h$ is weak,
one may approximately keep only the terms up to quadratic order in $\Delta A^h$ in the following expansion for the Lagrangian Eq.~(\ref{Lb}) as $L_b=\int d^2x\mathcal{L}_b$,
\begin{eqnarray}
\label{rewritingLb}
\begin{aligned}
\mathcal{L}_b[A^h]\approx \mathcal{L}_b[\bar{A}^h]&-iA^h_0n_v+ \frac{\partial\mathcal{L}_b}{\partial A^h_\alpha}[\bar{A}^h]\Delta A^h_\alpha\\&+\frac{1}{2}\frac{\partial^2\mathcal{L}_b}{\partial A^h_\alpha\partial A^h_\beta}[\bar{A}^h]\Delta A^h_\alpha\Delta A^h_\beta,
\end{aligned}
\end{eqnarray}
where $n_v=\sum_{\sigma,v}v b^\dagger_{\sigma,v}b_{\sigma,v}$ is the density of vorticity of spinon-vortices and the continuum limit is taken.
The current of vorticity $j^\alpha_s$, which can be divided into a paramagnetic part $j^\alpha_{s,\text{para}}$ and a diamagnetic part $j^\alpha_{s,\text{dia}}$, is manifest in Eq.~(\ref{rewritingLb}).
\begin{eqnarray}
\label{spinoncurrents}
\begin{aligned}
&j^\alpha_s[A^h]\equiv \frac{\partial \mathcal{L}_b}{\partial A^h_\alpha}[A^h]\equiv j^\alpha_{s,\text{para}}[\bar{A}^h]+j^\alpha_{s,\text{dia}}[A^h]\\
&j^\alpha_{s,\text{para}}[\bar{A}^h]\equiv \frac{\partial \mathcal{L}_b}{\partial A^h_\alpha}[\bar{A}^h]\\
&j^\alpha_{s,\text{dia}}[A^h]\approx\frac{\partial j^\alpha_s}{\partial A^h_\beta}[\bar{A}^h]\Delta A^h_\beta= \frac{\partial^2\mathcal{L}_b}{\partial A^h_\alpha\partial A^h_\beta}[\bar{A}^h]\Delta A^h_\beta.\\
\end{aligned}
\end{eqnarray}
Expanding the Gaussian integration
\begin{eqnarray}
\label{integratingoutspinons}
\begin{aligned}
e^{-S^\text{eff}_{\Delta A^h}}\equiv e^{-\int d\tau L^\text{eff}_{\Delta A^h}}\equiv\int\mathcal{D}b\;e^{-\int d\tau\int d^2x\mathcal{L}_b[A^h]}\\ 
\end{aligned}
\end{eqnarray}
up to quadratic order in $\Delta A^h$, we get
\begin{eqnarray}
\label{LeffAh}
\begin{aligned}
S^\text{eff}_{\Delta A^h}&=\int d\tau_1d\tau_2d^2x_1d^2x_2 
\frac{\chi_v}{2} A^h_0(\tau_1,x_1)A^h_0 (\tau_2,x_2)\\
&+\frac{K^{\alpha\beta}}{2}\Delta A^h_\alpha(\tau_1,x_1)\Delta A^h_\beta(\tau_2,x_2)\\
&-i\langle n_vj^\alpha_{s,\text{para}}\rangle A^h_0(\tau_1,x_1)\Delta A^h_\alpha(\tau_2,x_2).\\
\end{aligned}
\end{eqnarray}
where the spinon-vortex current-current correlation is defined as
\begin{eqnarray}
\label{spinonCCcorrelation}
\begin{aligned}
&K^{\alpha\beta}(\tau_1,\tau_2;x_1,x_2)\equiv K_\text{para}^{\alpha\beta}+K_\text{dia}^{\alpha\beta},\\
&K_\text{para}^{\alpha\beta}\equiv-\langle j^\alpha_{s,\text{para}}(\tau_1,x_1)j^\beta_{s,\text{para}}(\tau_2,x_2)\rangle,\\
&K_\text{dia}^{\alpha\beta}\equiv\left\langle\frac{\partial j^\alpha_{s,\text{dia}}(\tau_1,x_1)}{\partial A^h_\beta(\tau_2,x_2)}\right\rangle,
\end{aligned}
\end{eqnarray}
whose longitudinal components $K\equiv K^{xx}=K^{yy}$ is used in Eq.~(\ref{Aeffresult}).
The vortex density-density correlation is defined as
\begin{eqnarray}
\label{vortexvortexcorrelation}
\chi_v(\tau_1,\tau_2;x_1,x_2)\equiv\langle n_v(\tau_1,x_1)n_v(\tau_2,x_2)\rangle.
\end{eqnarray}
Explicit expressions and calculation of $K^{\alpha\beta}$ can be found in Appendix. 
After performing the Fourier transform and making use of translational invariance, we have
\begin{eqnarray}
\label{LeffAhFourier}
\begin{aligned}
S^\text{eff}_{\Delta A^h}&=T\sum_{\omega_n}\int \frac{d^2p}{(2\pi)^2}\frac{\chi_v}{2}A^h_0(\omega_n,p)A^h_0(-\omega_n,-p)\\
&+\frac{K^{\alpha\beta}}{2}\Delta A^h_\alpha(\omega_n,p)\Delta A^h_\beta(-\omega_n,-p)\\
&-i\langle n_vj^\alpha_{s,\text{para}}\rangle A^h_0(\omega_n,p)\Delta A^h_\alpha(-\omega_n,-p),
\end{aligned}
\end{eqnarray}
where $T$ is temperature, and $\omega_n$ is Matsubara frequency.
We then perform a derivative expansion for the response functions $\chi_v$, $K^{\alpha\beta}$, and $\langle n_vj^\alpha_{s,\text{para}}\rangle$. 
\begin{eqnarray}
\label{derivativeexpansion}
\begin{aligned}
\chi_v(&\omega_n,p)\approx \chi^s_ep^2+O(\omega_np^2,p^4)\\
K^{\alpha\beta}(&\omega_n,p)\approx[\chi^s_e\omega_n^2+\sigma^s_0\omega_n\text{sgn}(\omega_n)]\delta^{\alpha\beta}\\&+\chi^s_m(p^2\delta^{\alpha\beta}-p^\alpha p^\beta)+O(\omega_n^3,\omega_np^2,p^4)\\
\langle n_vj^\alpha_{s,\text{para}}&\rangle(\omega_n,p)\approx i[\chi^s_e\omega_n-\sigma^s_0\text{sgn}(\omega_n)]p^\alpha\\&+O(\omega^2_np,p^3),
\end{aligned}
\end{eqnarray}
where $\chi^s_e$ is the ``electric" susceptibility of the spinon-vortices, $\sigma^s_0$ is the DC conductivity of the spinon-vorticces, both of which were used in Eq.~(\ref{Aeffresult}), and $\chi^s_m<0$ is the ``magnetic" susceptibility of the spinon-vortices.
Equation (\ref{derivativeexpansion}) is the most general form based on rotational invariance (in the continuum limit), time-reversal invariance (when $\sigma^s_0=0$ in the superconducting phase) and current conservation (or gauge invariance).
The $\sigma^s_0$ terms explicitly break the time-reversal symmetry because finite conductivity requires dissipation.
The coefficients of the three quantities are related with each others by the continuity equation
\begin{eqnarray}
\label{continuityequation}
\begin{aligned}
i\partial_0 n_v+\partial_\alpha j^\alpha_{s,\text{para}}=0.
\end{aligned}
\end{eqnarray}
Substituting Eq.~(\ref{derivativeexpansion}) into Eq.~(\ref{LeffAhFourier}) and performing inverse Fourier transform, we are left with an effective Lagrangian density
\begin{eqnarray}
\label{Maxwell}
\begin{aligned}
\mathcal{L}^{\text{eff}}_{\Delta A^h} &= \frac{\chi^s_e}{2}(\partial_\alpha A^h_0-\partial_0\Delta A^h_\alpha)^2-\frac{\chi^s_m}{2}(\epsilon^{\alpha\beta}\partial_\alpha\Delta A^h_\beta)^2\\
&-i\sigma^s_0(\partial_\alpha A^h_0-\frac{1}{2}\partial_0\Delta A^h_\alpha)\Delta A^{h,\alpha}\\
&\equiv\mathcal{L}_\text{Maxwell}+\mathcal{L}_\text{cond}.
\end{aligned}
\end{eqnarray}
The Maxwell term describes linear ``dielectrics" and linear ``diamagnetism" of the spinon-vortices with respect to internal gauge field $\Delta A^h_\mu$.
The dissipative conductance term is only present above $T_c$.

After integrating out spinons $b$ and getting $L^\text{eff}_{\Delta A^h}$, we now integrating out $\Delta A^h$ to get an effective action $L^{\text{eff}}_{A^s}$ for gauge fields $A^s$.
In Eq.~(\ref{integratingout}), this is from the third line to the fourth line.
\begin{eqnarray}
\label{integratingoutAh}
\begin{aligned}
&e^{-S^\text{eff}_{A^s}}\equiv e^{-\int d\tau L^\text{eff}_{A^s}}\equiv\int\mathcal{D}\Delta A^h\;e^{-\int d\tau L_\text{MCS}+L^\text{eff}_{ \Delta A^h}}\\
&S^\text{eff}_{A^s}=T\sum_{\omega_n}\int \frac{d^2p}{(2\pi)^2} \frac{-1}{2\pi^2\chi^s_m}A^s_0(\omega_n,p)A^{s}_0(-\omega_n,-p)\\
&\;\;+\frac{1}{2\pi^2}\frac{\omega_n}{\chi^s_e\omega_n+\sigma^s_0\text{sgn}(\omega_n)}A^s_\alpha(\omega_n,p)A^{s,\alpha}(-\omega_n,-p).
\end{aligned}
\end{eqnarray}
When $\sigma^s_0=0$ in the superconducting phase, the second term in Eq.~(\ref{integratingoutAh}) reduces to a mass term just like the first term.
This is a well-known result that integrating out a Maxwell field with a mutual Chern-Simons (BF) coupling to another gauge field  will lead to a mass term of the coupled gauge field~\cite{SCisTO}.
Here, this mass gap is simply a reflection of the spinon-vortex gap.
When $\sigma^s_0\ne0$ in the normal state SVP phase, the gauge field $A^s_\alpha$ becomes gapless due to the de-confined spinon-vortices.

To integrate out the $A^s$ gauge field, we may first use the condition that the holons are condensed to
simplify $L^h$ by expressing it using the Higgs modes $H_h$ and the Goldstone modes $\theta$ according to Eq. (\ref{Hh}). Then using the unitary gauge to eliminate the Goldstone mode in the fifth and sixth lines of Eq.~(\ref{integratingout}). 
Substituting Eq.~(\ref{Hh}) to $L^h$ in Eq.~(\ref{Lagrangian_phasestring}) and taking the continuum limit, up to quadratic orders in $H_h$ and $\theta$, one obtains
\begin{eqnarray}
\label{Lagrangian_Higgs}
\begin{aligned}
\mathcal{L}_h=
 &i(\rho_h^0+2H_h\sqrt{\rho_h^0})(\partial_0\theta-A_0^e-A^s_0)+4\rho_h^0 UH_h^2\\
&+\frac{\rho_h^0+2H_h\sqrt{\rho_h^0}}{2m_h}(\partial_\alpha\theta-A^e_\alpha-A^s_\alpha)^2,
\end{aligned}
\end{eqnarray}
where $m_h=\frac{1}{2t_h}$ and $U=4t_h$~\cite{HardCoreBoson}, 
Taking the unitary gauge $\partial_\mu\theta-A_\mu\rightarrow -A_\mu$ to eliminate the Goldstone mode as usual, one then arrives at
\begin{eqnarray}
\label{unitarygauge}
\begin{aligned}
\mathcal{L}_h=&-i(\rho_h^0+2H_h\sqrt{\rho_h^0})(A_0^e+A^s_0)+4\rho_h^0 UH_h^2\\&+\frac{\rho_h^0+2H_h\sqrt{\rho_h^0}}{2m_h}(A^e_\alpha+A^s_\alpha)^2.
\end{aligned}
\end{eqnarray}
The term in the second line of Eq.~(\ref{unitarygauge}) contains a London term (sometimes called Meissner term or mass term), Eq.~(\ref{Meissner}), and a coupling term between the Higgs mode and electromagnetic fields, Eq.~(\ref{THGcoupling}).
On may define the bare superfluid stiffness $\rho_h^\text{bare}\equiv\frac{\rho_h^0}{m_h}$, and the THG coupling constant $k_\text{THG}\equiv\frac{2\sqrt{\rho_h^0}}{m_h}$.
The absence of a gapless Goldstone mode $\theta$ and the emergence of a gauge mass term in Eq.~(\ref{unitarygauge}) is the Anderson-Higgs mechanism ~\cite{Anderson_Higgs,HiggsMechanism}.

We finally integrate out the gauge fields $A^s_\mu$ to get an effective Lagrangian of the Higgs mode from the sixth to seventh line 
in Eq.~(\ref{integratingout}). The Gaussian integral can be evaluated by setting the integration variable equal to the point where the argument of the exponential is stationary~\cite{Weinberg}, i.e., solving the equations of motion of the gauge field $A^s_\mu$ and then substituting the solutions back to the Lagrangian. The equation of motion of $A^s_\alpha(\omega_n>0,p)$ is
\begin{eqnarray}
\label{AsEquationsofMotion}
\begin{aligned}
\frac{-1}{\pi^2}&\frac{\omega_n}{\chi^s_e\omega_n+\sigma^s_0}A^s_\alpha= (\rho_h^\text{bare}+H_hk_\text{THG})(A^e_\alpha+A^s_\alpha)\\
\end{aligned}
\end{eqnarray}
We are going to solve the above equation perturbatively by expanding it in powers of the Higgs modes $H_h$.
The zeroth order ($H_h$-independent) solution is
\begin{eqnarray}
\label{zerothorderAeff}
\begin{aligned}
(A^s_\alpha+A^e_\alpha)^{(0)} = \frac{\omega_n^2}{\omega_n^2+\pi^2\rho_h^\text{bare}(\chi^s_e\omega_n^2+\sigma^s_0\omega_n)}A^e_\alpha,
\end{aligned}
\end{eqnarray}
which exactly reduces to Eq.~(\ref{Aeffresult}) after analytic continuation.
The first order terms satisfy
\begin{eqnarray}
\label{firstorderAeff}
\begin{aligned}
\frac{1}{\pi^2}\frac{\omega_n}{\chi^s_e\omega_n+\sigma^s_0}A^{s,(1)}_\alpha+& \rho_h^\text{bare}(A^e_\alpha+A^s_\alpha)^{(1)}\\&=-H_hk_\text{THG}(A^e_\alpha+A^s_\alpha)^{(0)}.\\
\end{aligned}
\end{eqnarray}
The equation of motion of $A^s_0$ is
\begin{eqnarray}
\label{As0EquationsofMotion}
A^s_0 = -i\pi^2\chi^s_m(\rho_h^0+2H_h\sqrt{\rho_h^0}).
\end{eqnarray}
Substituting Eq.~(\ref{zerothorderAeff})-(\ref{As0EquationsofMotion}) into Eq.~(\ref{integratingoutAh}) and Eq.~(\ref{unitarygauge}) and under the temporal gauge $A^e_0=0$, we get
\begin{eqnarray}
\label{integratingoutAs}
\begin{aligned}
e^{-S^\text{eff}_{H_h}}\equiv& e^{-\int d\tau L^\text{eff}_{H_h}}\equiv\int\mathcal{D} A^s\;e^{-\int d\tau L_h[H_h,A^s;A^e]+L^\text{eff}_{A^s}}\\
S^\text{eff}_{H_h} =& \frac{1}{\beta}\sum_{\omega_n}\int\frac{d^2p}{(2\pi)^2} 4\rho_h^0 (U-\frac{1}{2}\pi^2\chi^s_m)H_h^2\\
&+\frac{1}{2}k_\text{THG}H_h[(A^e_\alpha+A^s_\alpha)^{(0)}]^2\\
&+\frac{1}{2}\rho^\text{bare}_h\frac{\omega_n^2}{\omega_n^2+\pi^2\rho_h^\text{bare}(\chi^s_e\omega_n^2+\sigma^s_0\omega_n)}(A^e_\alpha)^2.
\end{aligned}
\end{eqnarray}
This is the final effective action we are after.
From Eq.~(\ref{integratingoutAs}), we found the THG holon current $J^\alpha_{\text{THG}}=k_\text{THG}H_h(A^{e,\alpha}+A^{s,\alpha})^{(0)}$, which reproduces Eq.~(\ref{THGcuurent}), and the FHG holon current $J^\alpha=\rho_h^\text{bare}(A^{e,\alpha}+A^{s,\alpha})^{(0)}$, which reproduces Eq.~(\ref{LondonEquation}).
We remark that the FHG current defines the renormalized superfluid density $\tilde{\rho}_h\equiv J^\alpha/A^{e,\alpha}$ according to the London equation~\cite{Londonequation}, which exactly reproduces the results in Ref.~\cite{Han_SCstiffness}.
The equation of motion of the Higgs modes (in the low-energy-long-wavelength limit) is 
\begin{eqnarray}
\label{HiggsEOMfieldtheory}
\begin{aligned}
H_h=\frac{k_\text{THG}}{8\rho_h^0(2U-\pi^2\chi^s_m)} [(A^e_\alpha+A^s_\alpha)^{(0)}]^2, 
\end{aligned}
\end{eqnarray}
which reproduces Eq.~(\ref{HiggsEOM}) under the transverse gauge.

\subsection{C. Results}
With a complete mathematical formulation, we now calculate the phase shift of the THG signal defined by Eq.~(\ref{phaseofTHG})-(\ref{phaseshift}) and the renormalized THG intensity $I_{(3\omega)}/I^3_{(\omega)}$, where $I_{(3\omega)}\equiv\left\vert J_\text{THG}^\alpha\right\vert^2/(A^e_\alpha)^6$ and $I_{(\omega)}\equiv\left\vert J^\alpha\right\vert^2/(A^e_\alpha)^2$ are the raw intensities.
The phase shift and the renormalized THG intensity are directly measurable in the THG experiments, while the raw intensities are not since the effective electric field $A^e_\alpha$ inside the sample is affected by the reflectivity of the sample and other irrelevant details~\cite{NanlinWang_experiments}.

The results are calculated numerically using the following parameters. The superexchange coupling $J=120\text{meV}$. 
The effective superexchange coupling $J_\text{eff}$ is calculated according to Ref.~\cite{Phasestring_Ma_2014} by taking the backflow spinon $a_\sigma$ into account.
The holon hopping integral $t_h=3J$.
The frequency of the external gauge field $\hbar\omega_0=3\text{meV}\sim0.7 \text{THz}$, which is the same as that in Ref.~\cite{phaseshift_experiment}.
The spinon current-current correlation $K(\omega)$ is calculated on a $20\times20$ lattice with periodic boundary condition.
The DC conductivity of spinon-vortices $\sigma^s_0$ is got by introducing a small broadening $\eta$ of the spinon spectrum.
Our theory involves very few adjustable parameters.

\begin{figure}
	 \setlength{\belowcaptionskip}{-0.5cm}
	\vspace{-0.1cm}
	\begin{center}
		\includegraphics[width=8.5cm]{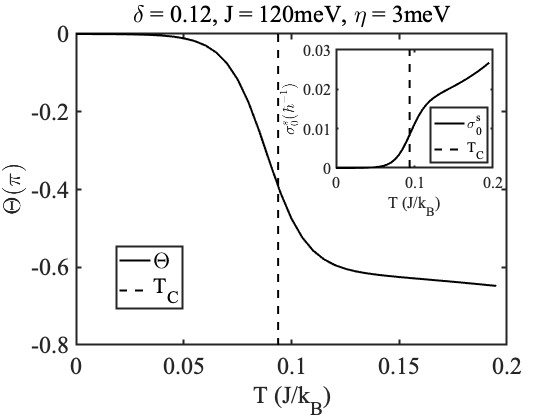}
	\caption{The phase shift of the THG signal calculated from the phase string theory. The phase shift happens at $T_c$, where the thermal spinon-vortices undergo an insulator-plasma phase transition (conductivity of spinons, $\sigma^s_0$, changes from zero to non-zero as shown in the inset). $\sigma^s_0$ is calculated by introducing a small broadening, $\eta$, of the spinon spectrum.}
	\label{fig_phaseshift}
	\end{center}
\end{figure}

As expected, in Fig.~\ref{fig_phaseshift}, there is a phase shift at $T_c$.
This phase shift of the THG signal is resulted from the phase shift of the gauge field $A^e_\alpha+A^s_\alpha$, which is, according to Eq.~(\ref{Aeffresult}) and Eq.~(\ref{zerothorderAeff}), directly related a jump in the conductivity of spinons, $\sigma^s_0$ across $T_c$, as shown in the inset of the figure.
This is consistent with the insulator-plasma phase transition of thermal spinon-vortices at $T_c$.
The phase shift is smaller than $\pi$ since the dissipation of vortex-like excitations is chosen to be finite.
The phase shift approaches $\pi$ in the strong dissipation limit, $\sigma^s_0(T>T_c)\propto\eta\gg\omega$, where $\eta$ is a broadening of the spinon spectrum.
The reason why $\sigma^s_0$ represented the dissipation can be understood from the fact that the electric resistivity $\rho=\pi^2\sigma^s_0$ when holons are condensed, according to Ref.~{\cite{Chen_resistivity}}.
The $T_c$ is determined using the formula $k_BT_c\approx6.4E_g$ given in~\cite{Mei_TC}, where $E_g$ is the gap of the spinon-rotons serving as a vortex fugacity.

\begin{figure}[t]
	 \setlength{\belowcaptionskip}{-0.5cm}
	\vspace{-0.1cm}
	\begin{center}
		\includegraphics[width=8.5cm]{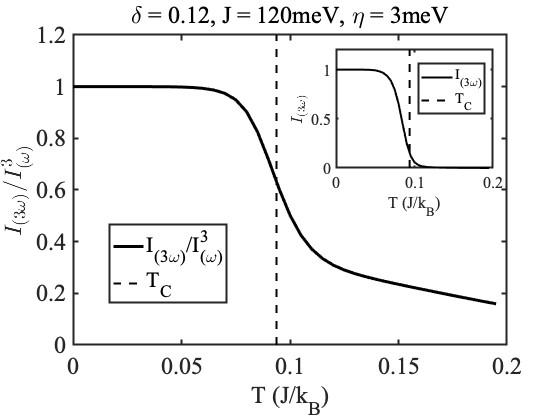}
	\caption{The THG intensity as a function of temperature in phase string theory. It decreases as increasing temperature, and has a jump a $T_c$. The unrenormalized one shown in the inset is often influenced by the weakening of the Meissner effect in real experiments.}
	\label{fig_intensity}
	\end{center}
\end{figure}

In Fig.~\ref{fig_intensity}, the renormalized THG intensity, in sharp contrast to conventional BCS superconductors, is still present above $T_c$ since the holons are still condensed.
The renormalized THG intensity decreases as increasing temperature, because $(A^e_\alpha+A^s_\alpha)/A^e_\alpha$ becomes smaller at higher temperature.
In other word, the screening of $A^e_\alpha$ by $A^s_\alpha$ is more effective at higher temperature, which reduces the effective THG coupling $k_{\text{THG}}^\text{eff}$.
The intensity undergoes a jump at $T_c$, since the screening of $A^e_\alpha$ by $A^s_\alpha$ becomes very strong above $T_c$.
(The screening above $T_c$ is complete at zero frequency and thus the superfluid density $\tilde{\rho}_h$ is driven to zero.)
The THG intensity will approach zero at a higher temperature $T_v$, above which the holons are no longer condensed.

After presenting our theoretical results, we now offer several remarks on the experimentally measured temperature dependence of THG intensity.
In Ref.~\cite{phaseshift_experiment}, in some temperature range inside $T<T_c$, the raw THG intensity $I_{(3\omega)}$ increases as increasing temperature.
This is supposed to be resulted from the weakening of the Meissner effect (decreasing screening of the driven field $A^e$), as was already noted in their paper.
Thus, to exclude this effect, one should study the renormalized intensity $I_{(3\omega)}/I^3_{(\omega)}$ rather than the raw one $I_{(3\omega)}$ shown in the inset of Fig.~\ref{fig_intensity}.
The renormalized intensity in our theory is consistent with the experimental ones in Ref.~\cite{phaseshift_experiment,NanlinWang_experiments}.
In Ref.~\cite{NanlinWang_experiments}, for the only one sample with the highest doping level, there is a peak feature in the temperature dependence of the THG intensity.
We think this feature might be resulted from the charge density fluctuations of the quasi-particles, which is not relevant to the Higgs mode.
It is thus beyond the scope of the present paper, so we did not include this contribution.

\section{IV. Conclusion and Discussion}
In this work, we have offered a unified understanding of the nonlinear THz THG response in the cuprate superconductor based on a microscopic framework of the $t$-$J$ model, in which the spin and charge degrees of freedom are intrinsically entangled by the phase string effect via the MCS gauge structure. In a conventional superconductor, the THG response can be uniquely attributed to the coupling between the EM field and the Higgs mode associated with the Cooper pair condensate below $T_c$. In the cuprate, the THG signal is found to persist above $T_c$ into the pseudogap regime. Furthermore, a $\pi$ phase shift has been observed across the SC transition. Within the MCS gauge theory, the Higgs field is the bosonic holon field, which experiences a condensation in both the SC phase and a lower pseudogap phase (i.e., the SVP/LPP) at underdoping. Consequently, the coupling between the EM field and the amplitude fluctuation (the Higgs mode) of the condensed holons can lead to a unique THG signal in the optical non-linear response measurement at either $T<T_c$ or $T>T_c$. Especially, a phase shift of the THG signal across $T_c$ or that of the corresponding EM-induced current response, can be understood by the confinement-deconfinement transition of spinon-vortices across $T_c$ in the MCS theory.

The THG experiments provide interesting evidence supporting the existence of a Higgs mode in the cuprate. Its significance lies in the fact that the nonlinear THG signal is not only unique, but also consistent with many other experimental measurements for the SC and SVP/LPP, which are unified by the same framework of the doped Mott insulator.  
For example, the universal relation between $T_c$ and the resonance-like spin gap ~\cite{Mei_TC}, the doping dependence of the superfluid density~\cite{Han_SCstiffness}, the Nernst effect~\cite{LPP,Phasestring_Nernst}, the transport experiments~\cite{Chen_resistivity,Song_transport} and many others~\cite{Phasestring_2007Review}.
By a unified framework, the understandings of different experiments can corroborate each other, e.g.,
the presence of the Nernst effect above $T_c$~\cite{Nernsteffect} is evidence for persistent holon condensation, which further indicates the existence of a Higgs mode above $T_c$.

The present work also provides a low-energy effective theory for future studies. Just like the Ginzburg-Landau theory (Abelian Higgs Model) can be viewed as an effective theory describing the Higgs mechanism and the Higgs modes dynamics in conventional superconductors~\cite{SCisTO}, our unified framework provides a generalized Gingburg-Landau theory~\cite{GeneralizedGLtheory,LPP} in cuprates, which covers both the SC phase and the lower pseudogap phase.

\section{Acknowledgments}
We thank Nan-Lin Wang, Ji-Si Xu, Zhi-Long Wang, Jing-Yu Zhao, and Zeyu Han for useful discussion.
The financial support by MOST of China (Grant No. 2021YFA1402101) and NSF of China (Grant No. 12347107) is acknowledged.

\appendix 
\section{Appendix. Diagonalizing $L_b$ and Calculating Current-Current Correlation of Spinon-Vortices}
In this appendix, we diagonalize the spinon-vortex Lagrangian Eq.~(\ref{Lb}) under mean-field approximation, and calculate the spinon-vortex current-current correlation function Eq.~(\ref{spinonCCcorrelation}).
We consider the cases of zero external magnetic field so that there is no Zeeman effect.
In these cases, the mean-field approximation is $A^h_0=\bar{A}^h_0=0, \epsilon^{\alpha\beta}\partial_\alpha A^h_\beta=\epsilon^{\alpha\beta}\partial_\alpha \bar{A}^h_\beta=\delta\pi$.
The RVB pairing mean-field parameter $\Delta^s$ and the spinon chemical potential $\lambda_b$ are obtained self-consistently in the following manner.
\begin{eqnarray}
\label{selfconsistentconditions}
\begin{aligned}
\Delta_s&=\sum_{\sigma,v}\langle b^{\dagger}_{i+\alpha,\sigma,v}b^{\dagger}_{i,\bar{\sigma},\bar{v}}e^{iv A_\alpha^h(i)}\rangle\\
1&=\sum_{\sigma,v}\langle b^\dagger_{i,\sigma,v}b_{i,\sigma,v}\rangle\\
0 &= \sum_{\sigma,v}v\langle b^\dagger_{i,\sigma,v}b_{i,\sigma,v} \rangle, 
\end{aligned}
\end{eqnarray}
The first line of Eq.~(\ref{selfconsistentconditions}) defines the pairing strength.
The second line means there is one spinon per site.
(The doping effect to the total number of local spins are taken into account via introducing a new kind of backflow spinon $a_{\bar{\sigma}}$ and keeping the number of spinon $b_\sigma$ fixed at one per site.
The spinons $a_{\bar{\sigma}}$ are high energy degrees of freedom, whose effect to the present study is simply renormalizing the coupling constant $J_\text{eff}$~\cite{Phasestring_review}.)
The zero-total-vorticity condition (in the absence of external flux) in the third line of Eq.~(\ref{selfconsistentconditions}) is required by the holon condensate.

The quadratic Lagrangian Eq.~(\ref{Lb}) can be diagonalized via Bogoliubov transformation
\begin{eqnarray}
\label{BogoliubovTrans}
\begin{aligned}
b_{i,\sigma,v}\equiv\sum_m w_{m,\sigma,v}(i)( u_{m}\gamma_{m,\sigma,v}-v_m\gamma_{m,\bar{\sigma},\bar{v}}^\dagger),
\end{aligned}
\end{eqnarray}
where $u_m$ and $v_m$ are taken to be real and satisfy $u_m^2-v_m^2=1$ to ensure bosonic commutation relations between $\gamma$ operators.
$w_{m,\sigma,v}(i)$ is one-spinon-vortex wavefunction, which is normalized by
\begin{eqnarray}
\label{wNormalization}
\begin{aligned}
\sum_m w_{m,\sigma,v}(i)w^*_{m,\sigma,v}(j)=\delta^i_{j}.
\end{aligned}
\end{eqnarray}
The requirement that $\gamma$ operators should diagonalize Eq.~(\ref{Lb}) reduces to the eigen-equation for $w_{m,\sigma,v}(i)$,
\begin{eqnarray}
\label{eigenequation}
\begin{aligned}
\xi_m w_{m,\sigma,v}(i)=-\frac{J_\text{eff}}{2}\sum_\alpha \Delta^s e^{-iv A^h_\alpha(i)}w_{m,\bar{\sigma},\bar{v}}^*(i+\alpha),
\end{aligned}
\end{eqnarray}
where $\xi_m$ is the eigenvalue, and the coefficients $u_m$ and $v_m$ are solved by
\begin{eqnarray}
\label{umvm}
\begin{aligned}
u_m &= \frac{1}{\sqrt{2}}\left(\frac{\lambda_b}{E_m}+1\right)^{\frac{1}{2}}\\
v_m &= \frac{1}{\sqrt{2}}\left(\frac{\lambda_b}{E_m}-1\right)^{\frac{1}{2}}\text{sgn}(\xi_m).
\end{aligned}
\end{eqnarray}
where $E_m=\sqrt{\lambda_b^2-\xi_m^2}$.
For simplicity, we shall consider the solution of a real pairing order parameter $\Delta^s$ so that the one-spinon-vortex wavefunction satisfy $w_{m,\sigma,v}=w^*_{m,\bar{\sigma},\bar{v}}$.
After this transformation, the Lagrangian Eq.~(\ref{Lb}) is diagonalized as
\begin{eqnarray}
\label{diagnolizedLb}
\begin{aligned}
L_b = \sum_{m,\sigma,v}\gamma^\dagger_{m,\sigma,v}(\partial_\tau+E_m)\gamma_{m,\sigma,v},
\end{aligned}
\end{eqnarray}
The mean-field parameters $\Delta_s,\lambda_b$, the eigen-energy $E_m$ and the wavefunction $w_{m,\sigma,v}$ can be calculated numerically by solving Eq.~(\ref{selfconsistentconditions}).

Next, we calculate the current-current correlation of the spinon-vortex.
According to Eq.~(\ref{spinoncurrents}), the currents are defined by
\begin{eqnarray}
\label{spinoncurrents_explicit}
\begin{aligned}
j^\alpha_s(i)&\equiv \frac{\partial\mathcal{L}_b}{\partial A^h_\alpha(i)}\equiv j^\alpha_{s,\text{para}}+j^\alpha_{s,\text{dia}}\\
&=-\frac{J_\text{eff}\Delta_s}{2}\sum_{\sigma,v}ivb^\dagger_{i+\alpha ,\sigma,v}b^\dagger_{i,\bar{\sigma},\bar{v}}e^{ivA^h_\alpha(i)}+h.c.\\
j^\alpha_{s,\text{para}}&\equiv j^\alpha_s|_{A^h=\bar{A}^h}\\
j^\alpha_{s,\text{dia}}
&\approx\frac{J_\text{eff}\Delta_s}{2}\sum_{\sigma,v}b^\dagger_{i+\alpha ,\sigma,v}b^\dagger_{i,\bar{\sigma},\bar{v}}e^{iv\bar{A}^h_\alpha(i)}\Delta A^h_\alpha(i)+h.c.
\end{aligned}
\end{eqnarray}
The spinon-vortex current-current correlation Eq.~(\ref{spinonCCcorrelation}) can be calculated directly by substituting Eq.~(\ref{spinoncurrents_explicit}) and then performing the Bogoliubov transformation Eq.~(\ref{BogoliubovTrans}). 
The results are
\begin{eqnarray}
\label{Kdia}
\begin{aligned}
K^{\alpha\beta}_\text{dia}(\tau_1,\tau_2;x_1(i),x_2(i^\prime))=J_\text{eff}\Delta_s^2\delta^{\alpha}_{\beta}\delta(\tau_1-\tau_2)\delta_{i,i^\prime},
\end{aligned}
\end{eqnarray}
where we also used Eq.~(\ref{selfconsistentconditions}), and
\begin{eqnarray}
\label{Kpara}
\begin{aligned}
&K^{\alpha\beta}_\text{para}(\tau_1,\tau_2;x_1(i),x_2(i^\prime))=\frac{J^2_\text{eff}\Delta_s^2}{4}\sum_{m,m^\prime} C^{\alpha\beta}_{mm^\prime}(i,i^\prime)\\
&\times\{[1+n_\text{B}(E_m)+n_\text{B}(E_{m^\prime})](u_mu_{m^\prime}+v_mv_{m\prime})^2\\
&\times\left[e^{(E_m+E_{m^\prime})(\tau_1-\tau_2)}-e^{-(E_m+E_{m^\prime})(\tau_1-\tau_2)}\right]\\
&+[n_\text{B}(E_m)-n_\text{B}(E_{m^\prime})](u_mv_{m^\prime}+u_{m^\prime}v_m)^2\\
&\times\left[e^{(-E_m+E_{m^\prime})(\tau_1-\tau_2)}-e^{(E_m-E_{m^\prime})(\tau_1-\tau_2)}\right]\},
\end{aligned}
\end{eqnarray}
where 
\begin{eqnarray}
\label{Cmmprime}
\begin{aligned}
&C^{\alpha\beta}_{mm^\prime}(i,i^\prime)\equiv\sum_{\sigma,v}e^{iv\bar{A}^h_\alpha(i)}e^{iv\bar{A}^h_\beta(i^\prime)}\\
&\times w^*_{m,\sigma,v}(i)w_{m^\prime,\sigma,v}(i+\alpha)w^*_{m^\prime,\sigma,v}(i^\prime)w_{m,\sigma,v}(i^\prime+\beta)\\
&-e^{iv\bar{A}^h_\alpha(i)}e^{-iv\bar{A}^h_\beta(i^\prime)}\\
&\times w^*_{m,\sigma,v}(i)w_{m^\prime,\sigma,v}(i+\alpha)w^*_{m^\prime,\sigma,v}(i^\prime+\beta)w_{m,\sigma,v}(i^\prime),
\end{aligned}
\end{eqnarray}
and 
$n_\text{B}(E)=\frac{1}{e^{E/T}-1}$
is the Bose-Einstein distribution function.
The Fourier transform is defined as
\begin{eqnarray}
\label{FouriertransformofCCcorrelation}
\begin{aligned}
K^{\alpha\beta}(\omega_n,p)\equiv &\sum_{i}e^{-i \vec{p}\cdot[\vec{x}_{1}(i)-\vec{x}_2(i^\prime)]}\\ 
&\times\int_0^\beta d\tau_1 e^{i\omega_n(\tau_1-\tau_2)}K^{\alpha\beta}(\tau_1,\tau_2;x_1,x_2),
\end{aligned}
\end{eqnarray}
where we have used translational invariance.
We take $p\rightarrow0$ in the continuum limit.
\bibliography{HiggsMode}
\end{document}